\documentclass[letterpaper,12pt]{article}   
\usepackage{osajnl2} 
\usepackage[draft]{hyperref} 
\usepackage{stmaryrd}
\usepackage{color} 
\usepackage{amsmath}

\begin{document}

\title{A Translational Polarization Rotator}


\author{David T. Chuss,$^{1^*}$ Edward J. Wollack,$^{1}$ Giampaolo Pisano,$^2$ Sheridan Ackiss,$^3$ Kongpop U-Yen,$^1$ Ming wah Ng$^2$ }

\address{$^1$Observational Cosmology Laboratory, Code 665, NASA Goddard Space Flight Center, Greenbelt, MD 20771, USA}
\address{$^2$University of Manchester, Manchester, UK}
\address{$^3$Georgia Tech, Atlanta, GA, USA}
\address{$^*$Corresponding author: David.T.Chuss@nasa.gov}

\begin{abstract}
We explore a free-space polarization modulator in which a variable phase is introduced between the right- and left-handed circular
polarization components and used to rotate the linear polarization of the outgoing beam relative to that of the incoming beam. In this device,
the polarization states are separated by a circular polarizer that consists of a quarter-wave plate in combination with a wire grid. 
A movable mirror is positioned behind and parallel to the circular polarizer. As the polarizer-mirror distance is separated, an incident linear polarization will be rotated through an angle that is proportional to the introduced phase delay. We demonstrate a prototype device that modulates  Stokes $Q$ and $U$ over a 20\% bandwidth, from 77 to 94 GHz.

\end{abstract}


\maketitle 

\section{Introduction}
Polarization modulation is the systematic mapping of an incident polarization state into a new polarization state for subsequent demodulation and detection.  This technique is useful for polarimetric applications in which the polarization signal is significantly smaller than the unpolarized background signal. Relevant applications include polarization contrast imaging and astronomical polarimetry \cite{Brosseau}.  

It is desirable for the polarization modulator to vary the polarization state but not the total amount of polarization. That is, in terms of Stokes parameters, an ideal modulator is subject to the condition
\begin{equation}
Q^2+U^2+V^2= \mathrm{constant}.
\end{equation}
This condition corresponds to a modulator that does not change the total coherence of the signal and makes the problem of measuring the polarization experimentally cleaner.  Modulators that satisfy this condition can be represented by unitary Jones matrices. In the homomorphically-equivalent Mueller formalism, the matrix representations of such modulators are orthogonal ({\it i.e.} the inverse is equal to the transpose).  This restriction limits the non-trivial operations to either a physical rotation or an introduction of a phase delay between orthogonal polarization components.  The latter are represented by rotations on the Poincar\'{e} Sphere \cite{Tinbergen} in which the basis and magnitude of the phase delay determine the axis and magnitude of the rotation, respectively. 

\begin{figure}[htpb]
	\centering
	\includegraphics[width=6in]{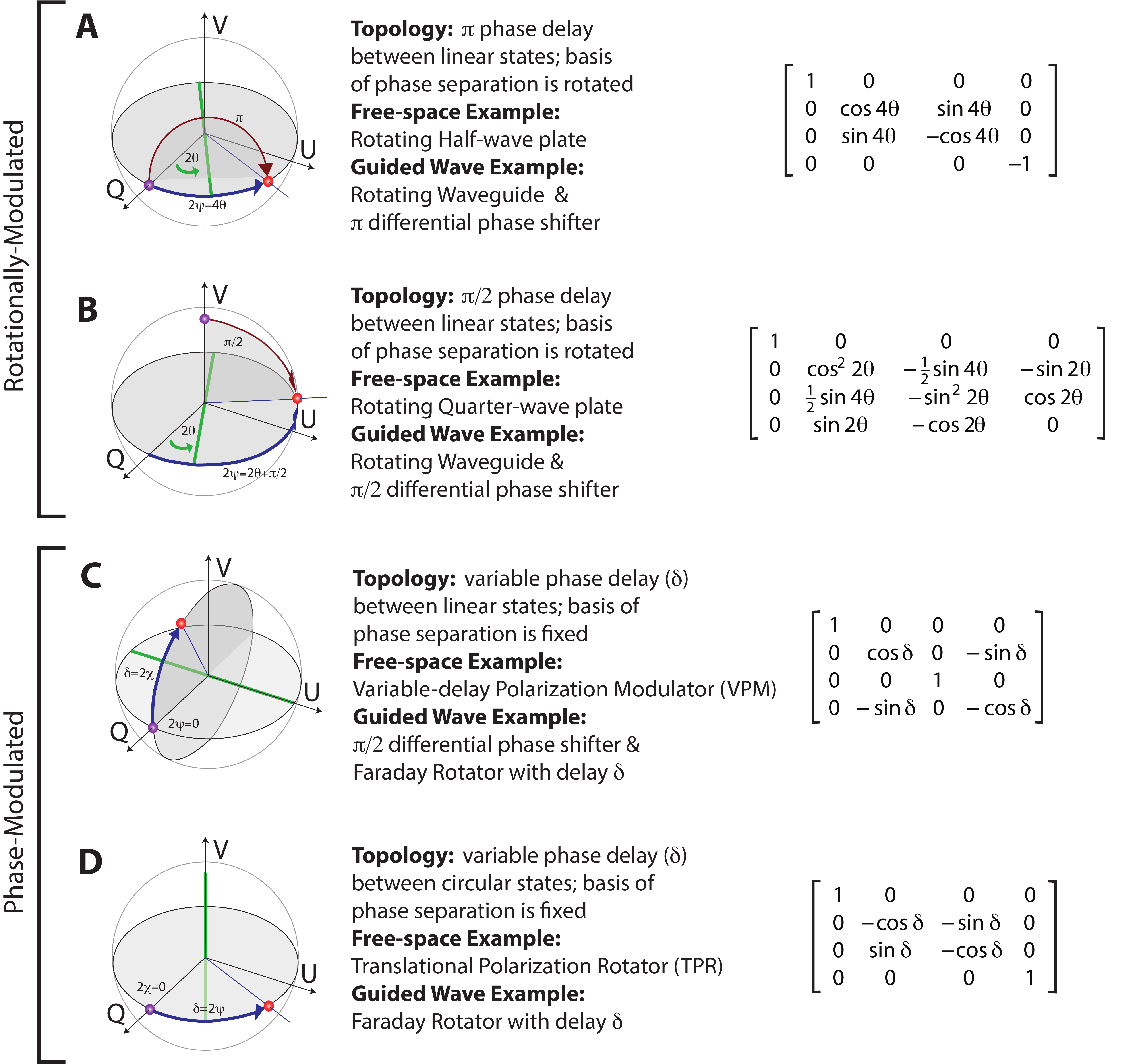}
	\caption{Schematic representations of ideal polarization modulators on the Poincar\'{e} Sphere are shown. The green line shows the axis connecting the two polarization states between which a phase delay, $\delta$, is introduced.  Blue arrows indicate the modulation path on the sphere. Red arrows indicate static phase delays. Deviations from ideal behavior are generally caused by finite bandwidth, biattenuance, differential reflection, {\it etc.} An example Mueller matrix for each architecture is shown on the right for free-space implementations ({\it i.e.}, A and B are in transmission; for C and D, the Mueller matrices are in reflection).}
	\label{fig:PS}
\end{figure}

Apart from the trivial example of instrument rotation, modulators typically vary either the basis of the system or the phase between two orthogonal polarizations. Figure~\ref{fig:PS} shows Poincar\'{e} sphere representations for a selection of unitary modulation schemes.  Corresponding Mueller matrices are also shown for specific implementations of each topology.
Figure~\ref{fig:PS}A shows a modulation scheme in which a constant phase delay of $\pi$ is introduced between two linear orthogonal polarizations. Modulation is accomplished by systematically changing the basis of phase separation in the $Q-U$ plane. A free-space realization of this topology is the rotating half-wave plate \cite{Hildebrand00, Savini06, Bryan10}. Topologically equivalent waveguide implementations are also possible \cite{Pisano11} that primarily differ in technological implementation and the dispersion relation. For ideal devices, it is possible to completely modulate linear polarization for a detector that is sensitive to linear polarization. 

Similarly, for measurement of circular polarization using a detector sensitive to linear polarization, it is possible to introduce a constant phase delay of $\pi/2$ between linear orthogonal polarizations and changing the basis of separation as shown in Figure~\ref{fig:PS}B.  Rotating quarter-wave plates and birefringent waveguides are examples of this architecture \cite{Pancharatnam55}.

Alternatively, as shown in Figure~\ref{fig:PS}C, it is possible to hold the polarization basis constant while introducing a variable phase delay between two orthogonal linear polarizations. A free-space example of this is the variable-delay polarization modulator (VPM). VPMs have been utilized to modulate polarization \cite{Chuss06,Krejny08, Chuss12} and have the potential to produce low and controllable systematic errors \cite{Chuss10} using small translational motions. This is a potential advantage for space flight applications as this concept can be realized with high reliability flexures that do not require rotational bearings. VPMs can only modulate a single linear Stokes parameter and therefore must rely on other degrees of freedom such as instrument rotation or separate optical paths in order to fully measure both $Q$ and $U$. For large telescopes, instrument rotation may lead to undesirable observational constraints, modulated instrumental polarization, or incomplete polarization coverage in the data. Thus it is desirable to seek a solution that fully modulates linear polarization in a single element to mitigate these concerns.

Introduction of a variable phase (Fig.~\ref{fig:PS}D) between the two \emph{circular} polarization states provides a means to fully modulate linear polarization. An example of this is implemented in a waveguide is a Faraday Rotator \cite{Nanos79, Ade09, Gault12} having a variable phase delay. In this case, the circular birefringence of ferrite material is altered as a function of applied magnetic field. 

In this paper, we present a concept for achieving a non-magnetic free-space modulator using topology `D'. We accomplish this by separating the two circular polarization states and introducing a variable, differential phase delay. We refer to this device as a Translational Polarization Rotator (TPR). Our implementation is related to the VPM architecture.  In Section~\ref{sec:TPR} we describe an implementation of a TPR. In Section~\ref{sec:measurement} we report laboratory results for a prototype TPR. We summarize in Section~\ref{sec:discussion}.


\section{The Translational Polarization Rotator}\label{sec:TPR}

\begin{figure}[htpb]
	\centering
	\includegraphics[width=5in]{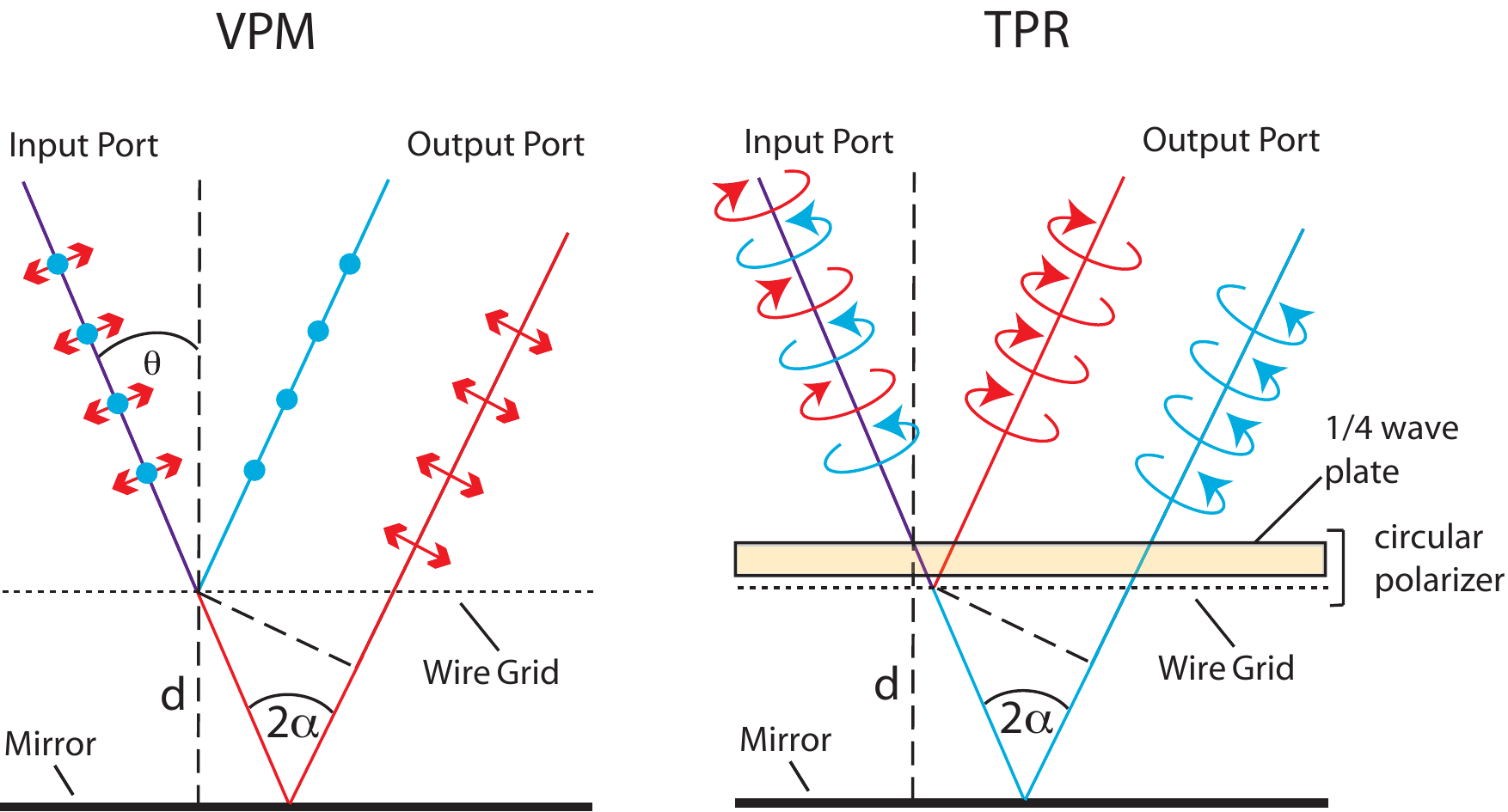}
	\caption{The topology for the TPR compared with the VPM is shown.  The TPR consists of a circular polarizer placed in front of and parallel to a movable mirror and introduces a variable phase delay between the orthogonal circular polarization states.}
	\label{fig:tpr}
\end{figure}

To realize a variable phase delay between right- and left-handed circular polarization components in a beam of radiation, we employ
the architecture illustrated on the right side of Figure~\ref{fig:tpr}.  This device consists of a quarter-wave plate placed in front of a VPM
with its fast (or slow) axis oriented at a 45$^\circ$ angle with respect to the VPM polarizer direction. The quarter wave-plate converts incoming
circular polarization states to orthogonal linear polarizations. The VPM then introduces a variable phase delay between the linear polarizations. As the
beam exits the device, another pass through the quarter-wave plate converts the linear polarizations back to right- and left- circular.

The TPR system can be analyzed using Jones matrices \cite{Jones41}. 
The simple analysis that follows is an ideal approach that is intended to illustrate the functionality of the TPR. The appropriateness of this approach relies on the absence of multiple coherent reflections or standing waves between the constituent elements of the TPR. Use of transfer matrices enables a more general treatment without this limitation \cite{Chuss12}.

The analysis of the system with the wave plate oriented at an arbitrary angle $\theta$ with respect to grid wires of a VPM is presented below.  The wave plate's phase delay and the angle between the wave plate and VPM can be set to specific values to realize the desired modulation.
The Jones matrix for the TPR can be expressed as
\begin{equation}
\overline{J}_{TPR}=\overline{J}_{WP}(-\theta,\beta)\overline{J}_{VPM}(0,\delta)\overline{J}_{WP}(\theta,\beta),
\end{equation}
where $\delta$ is the phase delay of the VPM, $\beta$ is the phase delay of the wave plate, and the Jones matrix for the wave plate is,
\begin{align}
\nonumber\overline{J}_{WP}(\theta,\beta)&=\overline{R}(-\theta)\overline{J}_{WP}(0,\beta)\overline{R}(\theta)\\
&=\left(\begin{array}{cc}
e^{i\frac{\beta}{2}}\cos^2{\theta}+e^{-i\frac{\beta}{2}}\sin^2\theta & -i\sin{2\theta}\sin{\frac{\beta}{2}} \\
 -i\sin{2\theta}\sin{\frac{\beta}{2}}& e^{i\beta/2}\sin^2{\theta}+e^{-i\frac{\beta}{2}}\cos^2\theta
 \end{array}\right).
 \label{eq:tpr1}
\end{align}
The origin of the angular coordinate system is the horizontal direction, parallel to the VPM wires. To derive this expression, the following definitions have been used,
\begin{align}
\overline{R}(\theta)=&
\left(\begin{array}{cc}
\cos{\theta} & \sin{\theta} \\
-\sin{\theta} & \cos{\theta}\end{array}\right)\\
\overline{J}_{WP}(0,\beta)=&
\left(\begin{array}{cc}
e^{i\frac{\beta}{2}} & 0 \\
0 & e^{-i\frac{\beta}{2}} \end{array}\right).
\end{align}
The angle $-\theta$ is negative in its first use in Equation~\ref{eq:tpr1} to account for the radiation passing through the wave plate twice, once in each direction.
The Jones matrix for the VPM is,
\begin{equation}
\overline{J}_{VPM}(0,\delta)=
\left(\begin{array}{cc}
e^{i\frac{\delta}{2}} & 0 \\
0 & -e^{-\frac{\delta}{2}} \end{array}\right),
\end{equation}
where the negative sign on the lower diagonal matrix element element arises from the parity change due to the reflection from the VPM.
Substituting appropriately, one finds the Jones matrix elements,
\begin{align}
\nonumber\overline{J}_{TPR}^{1,§1}&=
e^{i\frac{\delta}{2}}(e^{i\beta}\cos^4{\theta}+e^{-i\beta}\sin^4{\theta}+2\cos^2{\theta}\sin^2{\theta})-e^{-i\frac{\delta}{2}}\sin^2{2\theta}\sin^2{\frac{\beta}{2}}\\
\nonumber\overline{J}_{TPR}^{1,2}&=
i\sin{2\theta}\sin^2{\frac{\beta}{2}}\left[e^{i\frac{\delta}{2}}(e^{i\frac{\beta}{2}}\cos^2{\theta}+e^{-i\frac{\beta}{2}}\sin^2{\theta})
+e^{-i\frac{\delta}{2}}(e^{i\frac{\beta}{2}}\sin^2{\theta}+e^{-i\frac{\beta}{2}}\cos^2{\theta})\right]\\
\nonumber\overline{J}_{TPR}^{2,1}&=
-i\sin{2\theta}\sin^2{\frac{\beta}{2}}\left[e^{i\frac{\delta}{2}}(e^{i\frac{\beta}{2}}\cos^2{\theta}+e^{-i\frac{\beta}{2}}\sin^2{\theta})
+e^{-i\frac{\delta}{2}}(e^{i\frac{\beta}{2}}\sin^2{\theta}+e^{-i\frac{\beta}{2}}\cos^2{\theta})\right]\\
\overline{J}_{TPR}^{2,2}&=
-e^{-i\frac{\delta}{2}}(e^{-i\beta}\cos^4{\theta}+e^{i\beta}\sin^4{\theta}+2\cos^2{\theta}\sin^2{\theta})+e^{i\frac{\delta}{2}}\sin^2{2\theta}\sin^2{\frac{\beta}{2}}.
\end{align}
Setting $\theta=\pi/4$ and $\beta=\pi/2$, one finds the Jones matrix for a monochromatic TPR,
\begin{equation}
\overline{J}_{TPR}\left(\frac{\pi}{4},\frac{\pi}{2}\right)=
i\left(\begin{array}{cc}
\sin{\delta}& \cos{\delta} \\
-\cos{\delta}&\sin{\delta} \end{array}\right).
\end{equation}
The associated Mueller matrix can be identified by examining the transformation of the density matrix,
\begin{equation}
D^\prime= \overline{J}^\dagger_{TPR}D\overline{J}_{TPR} .
 \end{equation}
 $D$ and $D^\prime$ can each be decomposed in the Pauli basis \cite{Brosseau}:
\begin{align}
D&=I\overline{\sigma_0}+Q\overline{\sigma_1}+U\overline{\sigma_2}+V\overline{\sigma_3}\\
&=
I\left(\begin{array}{cc}1& 0\\0& 1\end{array}\right)+
Q\left(\begin{array}{cc}1& 0\\0& -1\end{array}\right)+
U\left(\begin{array}{cc}0& 1\\1& 0\end{array}\right)+
V\left(\begin{array}{cc}0& -i\\i& 0\end{array}\right).
\end{align}
From this, the transformation of the Stokes parameters can be determined and organized into the Mueller matrix \cite{Tinbergen} for the system,
\begin{equation}
\overline{M}_{TPR}\left(\frac{\pi}{4},\frac{\pi}{2}\right)=
\left(\begin{array}{cccc}
1&0&0&0\\
0&-\cos{\delta}&-\sin{\delta} &0\\
0&\sin{\delta} & -\cos{\delta} & 0\\
0 & 0 & 0 & 1
\end{array}\right).
\end{equation}
The linear Stokes parameters transformation,
\begin{align}
\nonumber Q^\prime&=-Q\cos{\delta}-U\sin{\delta}\\
U^\prime&=Q\sin{\delta}-U\cos{\delta},
\label{eq:TPR}
\end{align}
demonstrates that the TPR architecture does indeed inject a variable phase delay between left- and right- circular polarization.  

\section{Measurement}\label{sec:measurement}

\begin{figure}[htpb]
	\centering
	\includegraphics[width=6in]{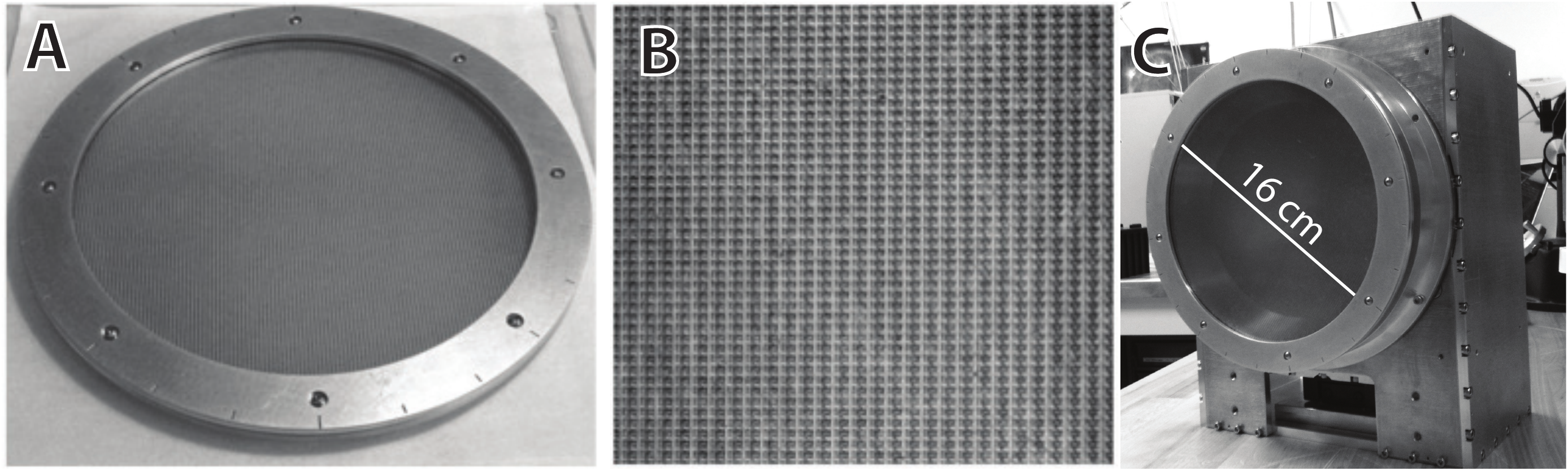}
	\caption{The metal-mesh wave plate is shown (A) along with a close-up view of its surface (B).  The prototype TPR is shown in (C).}
	\label{fig:tprphoto}
\end{figure}

A prototype TPR has been constructed using a metal mesh quarter-wave plate of the type presented in \cite{Pisano12, Maffei12}. This engineered birefringent dielectric device was mounted to the front of the Hertz VPM prototype \cite{Chuss12, Krejny08} (see Fig.~\ref{fig:tprphoto}), and the grid-mirror separation was controlled and measured using a manual linear micrometer stage. The metal mesh quarter-wave plate is based on photolithographic techniques used in the past to realize half-wave plates \cite{Pisano08}. This device provides a phase shift between orthogonal linear polarizations of $89.2^\circ\pm1.5^\circ$ over a 40\% bandwidth (75-110 GHz). The transmittance for the two polarizations is matched to 2\% from 77-94 GHz where the experimental efforts were concentrated. The biattenuance of the wave plate in this configuration leads to a finite modulated instrumental polarization. The bandwidth has been limited to control the magnitude effect.

\begin{figure}[htpb]
	\centering
	\includegraphics[width=6in]{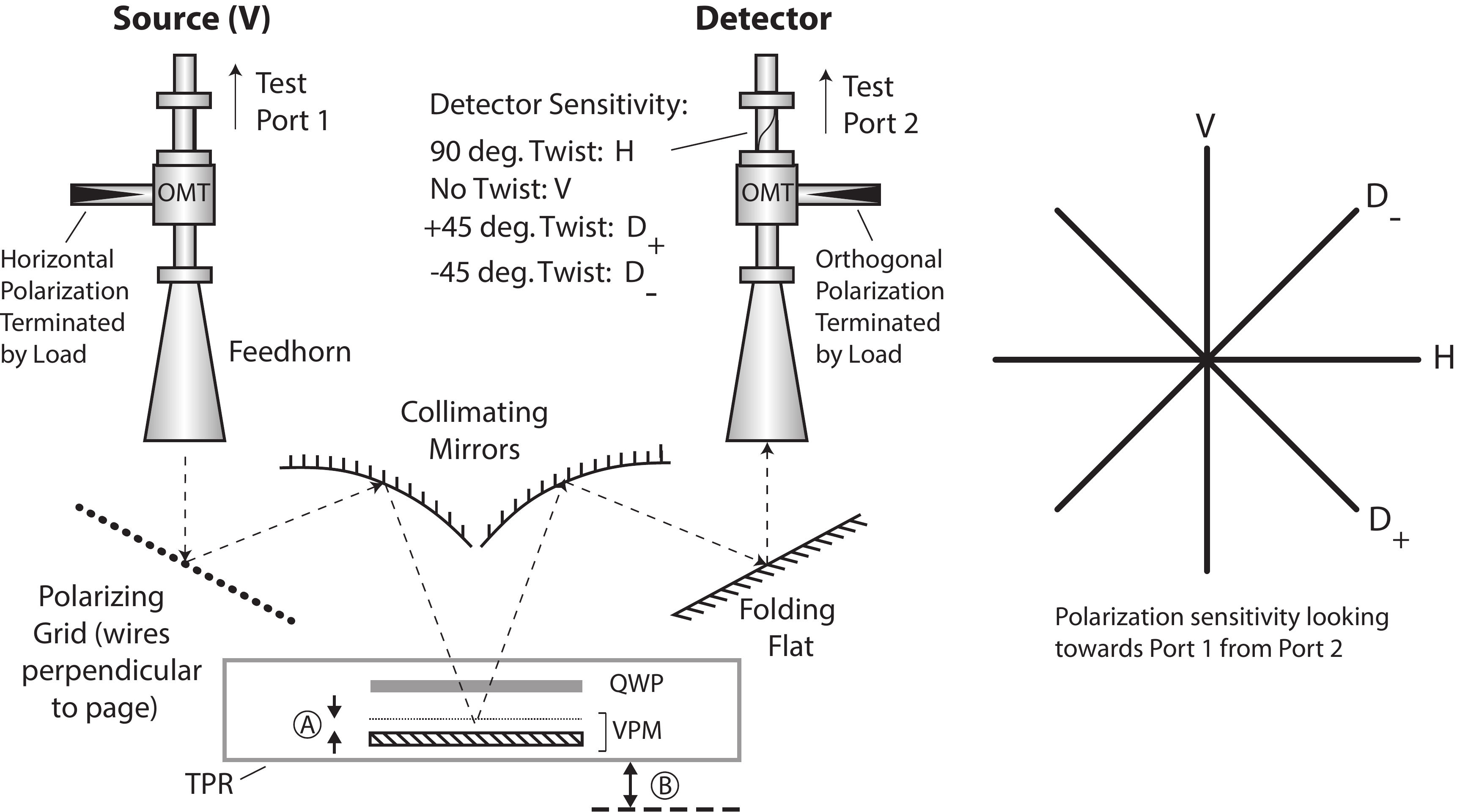}
	\caption{The test setup used to validate the VPM is shown (left). The grid-mirror separation is given by A, and the overall displacement of the TPR is given by B.  The rotational coordinate system used for the polarization measurements is shown on the right.}
	\label{fig:setup}
\end{figure}

To test the operation of the TPR, the prototype was included in the test setup shown in Figure~\ref{fig:setup}. This largely follows the experimental approach utilized in \cite{Chuss12}. The test setup used a pair of feed horns to
couple the quasioptical testbed to an Agilent PNA-X vector network analyzer.  Microwaves polarized in the vertical ($-U$) direction are transmitted from Port 1 (designated the ``Source'').   A polarizing wire grid was used to further define the polarization state and served to redirect the radiation to an ellipsoidal mirror that maps the feed's beam waist onto the TPR. A second, identical ellipsoidal mirror re-mapped the beam waist into a second feed horn (the ``Detector'') attached to Port 2 of the PNA-X. Orthomode Transducers (OMT) were used to terminate the unused polarization in each feed horn. The quarter-wave plate was tilted at an angle $>5^\circ$ with respect to the VPM wire grid, to limit the influence of trapped modes in the cavity between the wave plate and the grid.  
By reducing the distance between these elements, this effect can be moved out of the signal band.

A second linear micrometer stage was inserted between the TPR and the test setup.  This linear stage was used to vary the position of the TPR (``B'' in Figure~\ref{fig:setup}) relative to the rest of the optics. By taking measurements of the response at different positions of the TPR, it was possible to use the varied phase to separate the TPR response from that of the rest of the optics using a procedure similar to that outlined by Eimer et al. \cite{Eimer11}.  Thus each $S_{21}$ scattering parameter measurement described below is a composite of four measurements taken at 400 $\mu$m intervals for the TPR position.  

We characterized the polarization transfer function of the TPR by measuring the normalized linear Stokes parameters, $q\equiv Q/I$ and $u\equiv U/I$, at Port 2.  This was achieved by measuring the complex $S_{21}$ scattering parameter as a function of grid-mirror separation at four rotations of the horn attached to Port 2: 0$^\circ$ ($V$), 90$^\circ$ ($H$), 45$^\circ$ ($D_+$), and -45$^\circ$ ($D_-$).  These angles were measured at the feed horn flange and their relative error is roughly $\pm 0.2^\circ$.  To mechanically facilitate coupling to the millimeter-wave receiver module,  a short (1.5-inch) section of appropriate waveguide twist between was used  between Port 2 and the OMT for measurements of $H$, $D_-$, and $D_+$.  The Stokes parameters can be extracted from these measurements, 

\begin{equation}
q(d) = \frac{H(d) - f_u V(d)}{H(d) + f_u V(d)}
\end{equation}
\begin{equation}
u(d) = \frac{D_+(d) - f_q D_-(d)}{D_+(d) + f_q D_-(d)}.
\end{equation}

\begin{figure}[htpb]
	\centering
	\includegraphics[width=5.5in]{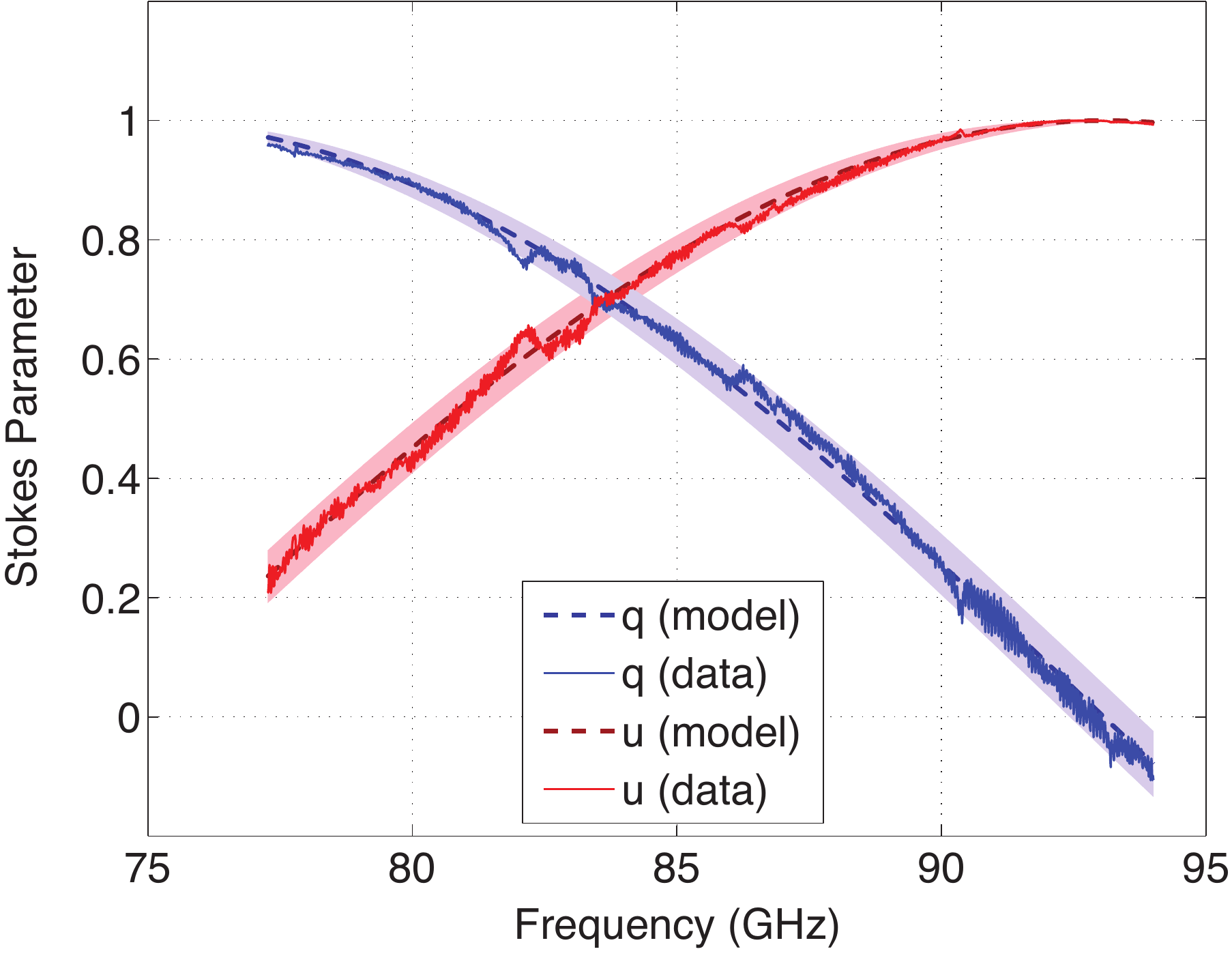}
	\caption{The measured $q$ and $u$ spectra are shown for a grid-mirror separation of 1143 $\mu$m. The dashed lines are the corresponding instrument models for $q$ and $u$. The shaded region corresponding to each of the Stokes parameters illustrates the 
effect of a $\pm$15 $\mu$m uncertainty in the grid-mirror separation. }
	\label{fig:example}
\end{figure}

The values $f_q$ and $f_u$ are the relative gain of the system between the different rotations. The gain can vary upon rotation due to changes in the feed illumination or waveguide twist ohmic loss.  The $D_+$ and $D_-$ measurements each employ 1.5 inch twists, so we set $f_u=1$.  The measurements of $V$ do not include
a twist section, while those for $H$ do.  Therefore, to account for the loss imbalance in the measurement of $V$ relative to that of $H$, $f_q=0.99$. An example of the calibrated $q$ and $u$ data is shown in Figure~\ref{fig:example}.

\begin{figure}[htpb]
	\centering
	\includegraphics[width=5.5in]{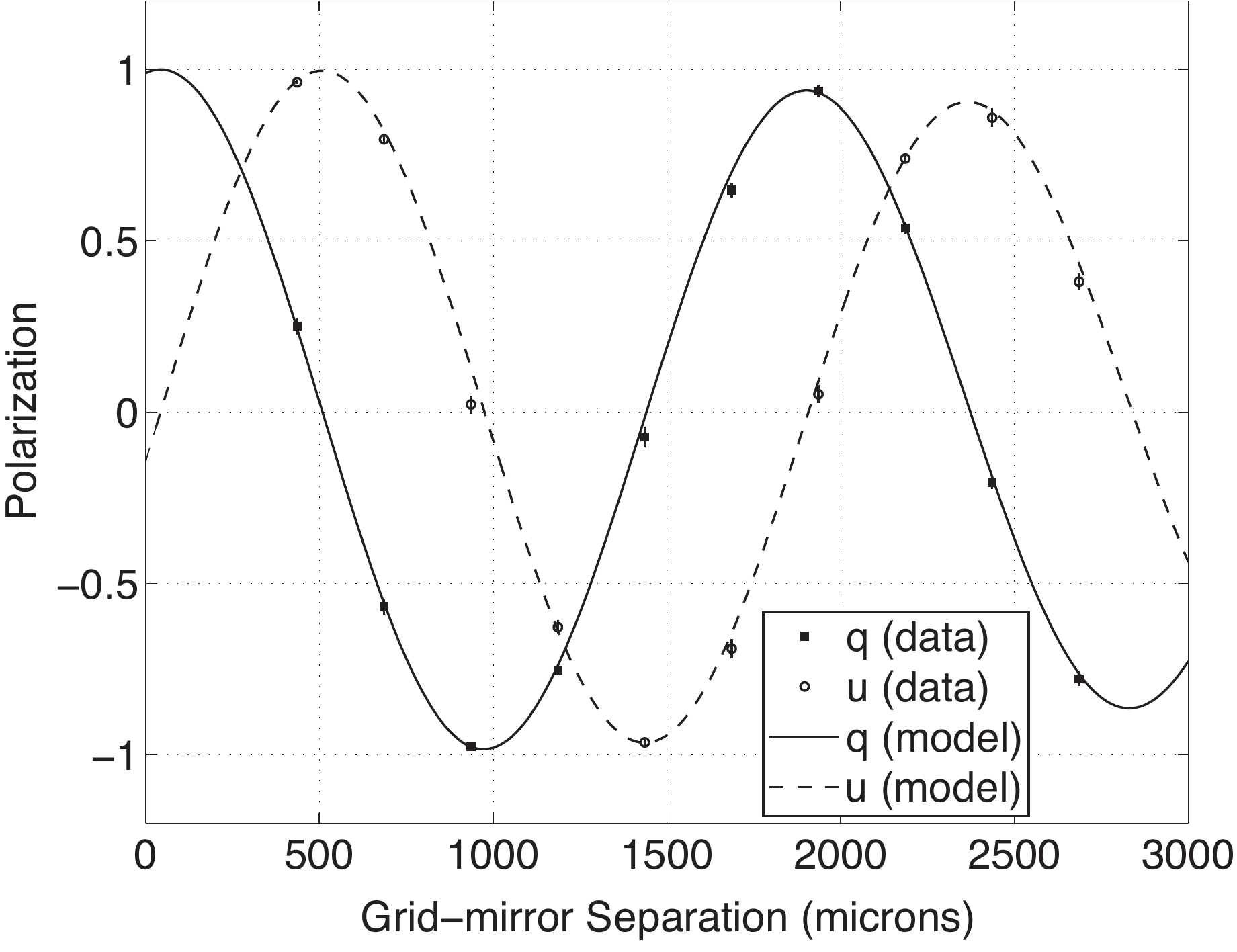}
	\caption{The response of the TPR to an incident $-q$ (vertically-polarized) signal is shown. The measured $q$ and $u$ response for the integrated 77-94 GHz response is superposed on the integrated response expected from theory.}
	\label{fig:response}
\end{figure}

The calibrated measurements have been integrated over the 77-94 GHz band and are shown in Figure~\ref{fig:response}. The model responses for $q$ and $u$ are similarly integrated,
\begin{align}
\nonumber q(\mathrm{model})&=\int_{\Delta\lambda}\cos{\delta(\lambda)}d\lambda\\
u(\mathrm{model})&=-\int_{\Delta\lambda}\sin{\delta(\lambda)}d\lambda.
\end{align}
These expressions come from Equation~\ref{eq:TPR} after setting $Q=-1$ and $U=0$ and integrating over the bandwidth, $\Delta\lambda$.  The single free parameter in this model is a constant offset in grid-mirror separation that we have chosen to minimize the variance between the model and the data. Because the wavelength is much greater the diameter of the wire of the polarizing grid, the phase can be approximated by $\delta(\lambda)\sim4\pi d \cos{\alpha}/\lambda$ where $d$ is the grid-mirror separation, $\alpha=20^\circ$ is the incidence angle of the radiation on the modulator, and $\lambda$ is the wavelength \cite{Chuss12}. 

The reported vertical error bars at each grid-mirror separation in Figure~\ref{fig:response} are calculated from the instrument model and the observed frequency-dependent polarization response ({\it e.g.} see Fig.~\ref{fig:example}). The variance between these two quantities is minimized by marginalizing over the grid-mirror separation.  This uncertainty can be explained by a combination of the errors in the {\it in situ} calibration procedure described above and residual trapped modes between the quarter-wave plate and the VPM.  The error analysis described reveals that the dominant observed systematic error (2\% RMS for $q$, 3\% RMS for $u$) is consistent with a RMS grid-mirror separation uncertainty of $15\, \mu$m. The effect of this uncertainty in the polarization spectrum is shown as shaded bands in Figure~\ref{fig:example} and is more significant at zero crossings than at the extrema of the modulation functions.
Future implementations beyond this simple proof-of-concept can eliminate these artifacts. This can be achieved by directly metering the grid-mirror separation ({\it e.g.} with capacitive sensors, glass scales, {\it etc.}) and reducing the spacing between the quarter-wave plate and the grid. 

\section{Summary}\label{sec:discussion}
A prototype TPR has been constructed and validated using a vector network analyzer. Residual deviations between the model and data are presently dominated by the uncertainty in the grid-mirror position. Future implementations in which this is mitigated will enable more detailed investigation of the effects of gain uncertainty and biattenuance in the quarter-wave plate and will focus on implementing TPRs in astronomical polarimetry systems.  The characteristics of the circular polarizer will ultimately determine the TPR's performance as an element of a broadband polarimeter.  The demonstrated approach is potentially useful for astronomical polarimetry in the millimeter through far-infrared in that it enables full linear polarization modulation with a single reflective element.   

\section*{Acknowledgements}
This work was funded by an internal research award at GSFC.  Funding for S. Ackiss was provided in part through the Undergraduate Student Researchers' Program and the NASA Georgia Space Act Consortium.
\bibliographystyle{osajnl}

\end{document}